\documentclass{vgtc}                          %

\graphicspath{{figures/}{pictures/}{images/}{./}} %

\usepackage{times}                     %

\usepackage{tabularx}                  %
\usepackage{booktabs}                  %
\usepackage{lipsum}                    %
\usepackage{mwe}                       %
\usepackage{mathptmx}                  %
\usepackage{graphicx}
\usepackage{standalone}
\usepackage{tikz}
\usepackage{amssymb}
\usepackage{stfloats}
\usepackage{soul,xcolor}
\usetikzlibrary{shapes.geometric, arrows.meta, positioning, fit, backgrounds, calc}
\usetikzlibrary{arrows.meta, positioning, fit, backgrounds, calc}

\definecolor{PalBlack}{HTML}{000000}
\definecolor{PalOrange}{HTML}{E69F00}
\definecolor{PalSky}{HTML}{56B4E9}
\definecolor{PalTeal}{HTML}{139258}
\definecolor{PalYellow}{HTML}{F0E442}
\definecolor{PalBlue}{HTML}{0072B2}
\definecolor{PalRust}{HTML}{D55E00}

\definecolor{PalRQ}{HTML}{ffd553}
\definecolor{PalRQStroke}{HTML}{b47004}

\definecolor{PalIntegration}{HTML}{03809f} %

\definecolor{PalQual}{HTML}{894ca8}       %

\colorlet{SetupNavy}{PalBlack!40}
\colorlet{SetupLightBg}{PalBlack!1}   %
\colorlet{SetupBorder}{PalBlack!50}   %

\colorlet{QuantFill}{PalTeal!10}
\colorlet{QuantStroke}{PalTeal}

\colorlet{QualFill}{PalQual!14}
\colorlet{QualStroke}{PalQual}

\colorlet{IntegHead}{PalIntegration!95}                        %
\colorlet{AnalysisFill}{PalIntegration!10!PalBlue!10}                   %
\colorlet{AnalysisStroke}{PalIntegration!55!PalBlack}        %
\colorlet{AnalysisZoneBg}{PalIntegration!6}                  %

\colorlet{RqFill}{PalRQ!20}
\colorlet{RqStroke}{PalRQStroke}

\colorlet{SetupZoneBg}{PalBlack!4}
\colorlet{AnalysisZoneBg}{PalBlack!2}   

\definecolor{myblue}{HTML}{c3e4ff}   %
\definecolor{myorange}{HTML}{fedbba} %

\definecolor{inlinequote}{HTML}{FFFBED}
\sethlcolor{inlinequote}

\newcommand{\Dashboard}{%
  {\setlength{\fboxsep}{2pt}%
   \colorbox{myblue}{\textsf{\textbf{Dashboard}}}%
  }%
}

\newcommand{\Maker}{%
  {\setlength{\fboxsep}{2pt}%
   \colorbox{myorange}{\textsf{\textbf{Maker}}}%
  }%
}

\newcommand{\inquote}[1]{\hl{#1}}

\newcommand{\theme}[1]{%
  \vspace{.5em}\noindent{%
    \bfseries
    \fontsize{9}{11}\selectfont
    #1%
  }
}

\makeatletter
\renewenvironment{quote}
  {\list{}{\setlength{\leftmargin}{.6cm}%
           \setlength{\rightmargin}{.6cm}%
           \setlength{\parsep}{\z@}%
           \topsep=4.5pt      %
           \parsep=0pt      %
           \partopsep=1pt}  %
   \item\relax}
  {\endlist}
\makeatother

\newcommand{\qref}[1]{%
  {\hyperref[q:#1]{#1}}%
}

\newcommand{\orcid}[1]{\href{https://orcid.org/#1}{\includegraphics[width=10pt]{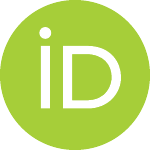}}}
\usepackage{xcolor}

\newcommand{\edit}[1]{\textcolor[HTML]{000000}{#1}}

\onlineid{0}
\vgtccategory{Research}
\vgtcinsertpkg

\title{
 Text-Based Visualizations for Learning Analytics: Perspectives from STEM Education Researchers
}

\author{Huyen N. Nguyen\orcid{0000-0001-6554-2327}\thanks{e-mail: huyen@lanl.gov}\\ %
        \scriptsize Los Alamos National Laboratory %
\and Kathleen A. Bowe\orcid{0000-0001-7992-2937}\thanks{e-mail: kathleen.jeffery7@gmail.com}\\%
     \scriptsize University of New Hampshire %
\and Minh-Huyen Nguyen\orcid{0009-0004-4687-1933}\thanks{e-mail: huyennm@soict.hust.edu.vn}\\ %
     \scriptsize Hanoi University of Science and Technology %
\and Kit Thompson\thanks{e-mail: kitty.alice.t@gmail.com}\\ %
     \scriptsize University of Washington Bothell
\and Caleb M. Trujillo\orcid{0000-0002-6190-0577}\thanks{e-mail: calebtru@uw.edu}\\ %
     \scriptsize University of Washington Bothell}

\abstract{
   Visualizing student reflections collected throughout a course can help instructors track changes in student language and inform teaching decisions. 
Interpreting these changes, however, requires tools that connect overall patterns with individual responses, and how to design such visualizations for learning analytics remains an open question. 
To address this gap, we conducted a mixed-methods study with 10 STEM education researchers who offered \textit{dual} perspectives as both instructors and qualitative researchers. 
We used two visualization platforms as probes to elicit expert perspectives: the Journal Data Dashboard, for analyzing formative assessments, and WordStream Maker, for authoring custom visualizations. 
Participants evaluated both platforms by analyzing weekly student journal responses from a data visualization course. 
Findings revealed shared prioritization of context and control, discipline-rooted differences in perceived usefulness, and, notably, \textit{epistemological} disagreement about frequency as viable evidence in qualitative work. 
Some participants regarded frequency-based visualization as a productive entry point to qualitative analysis, while others argued that it conflicts with the values of qualitative data analysis and may obscure rare but critical responses. 
These findings support the argument that perceived ease of use and usefulness, context of use or role, and epistemic frames are relevant to users of qualitative data visualization and learning analytics tools. 
We discuss design implications for future tools to better integrate quantitative techniques with qualitative interpretation. 
Transcripts, tutorials, and all supplementary materials are available at \url{https://osf.io/z2f8d}.

}

\nocopyrightspace

\begin{document}
\maketitle

\section{Introduction}
\label{sec:intro}

Data-driven learning analytics can reveal emerging patterns within a student cohort and provide concrete evidence for improving the learning environment. For example, tracking how students' language shifts over time can give instructors insight into where students are and how their thinking is developing. Much of the richest evidence for this purpose is in written form, such as reflections, open responses, and formative assessments. However, the qualitative and open-ended nature of this format makes it highly context-laden, requiring substantial time and effort to identify overarching trends without losing the nuances of individual contexts~\cite{tan2024epistemic}.

We use the term \textit{qualitative learning analytics} to describe this kind of analysis. Situated within broader efforts to use qualitative studies to provide richer context for the visualization community~\cite{losev2022embracing}, a growing body of work characterizes visualization for qualitative data exploration~\cite{chandrasegaran2017integrating, Interplay, meng2026designing} and, in particular, for educational settings~\cite{diehl2021visguided}. However, little of this work focuses specifically on qualitative learning analytics, which demands approaches that track the temporal evolution of student language while preserving the specific contexts in which that language appears.V

To design qualitative learning analytics tools, designers must account for how qualitative data are used in education and research. Visualizations should connect high-level trends with close reading, supporting instructors as they assess student work and provide feedback, and researchers as they analyze student writing and other textual data. 
Visualizing qualitative data may incur a risk in striping away important contexts. Aggregation and emphasis can conceal important information from a viewer, raising ethical concerns ~\cite{correll2019ethical} and oversimplifying rich qualitative data ~\cite{vis_literacy, aigner2011visualization}. These concerns parallel critical approaches to information visualization that critique \emph{whose} questions an encoding ultimately serves~\cite{dork2013critical}. 

Natural language processing~\cite{wang2016ideas} and, more recently, assistive AI tools~\cite{Davidson2024AtlasAI} are more recent approaches to support qualitative data visualization. As computational approaches they analyze, codify and display textual patterns. These approaches can scale and automate through pattern recognition. However, these approaches often run counter to the researcher-as-instrument view espoused by many qualitative researchers. Similarly, analyses of computationally-based text visualizations seldom assess trustworthiness or credibility~\cite{wanner2014state, kucher2022interdisciplinary}, two major values of qualitative researchers. Therefore, textual learning analytics tools are more likely to be accepted if they scale computational analysis while preserving the unique benefits of qualitative approaches.

Despite their demonstrated utility, we still lack a broad understanding of how well these tools fit educational qualitative workflows, especially for people who both design analyses and use them in their own teaching. \edit{\textbf{Instructors who are also qualitative education researchers} sit at this intersection:} they bring disciplinary expectations about qualitative rigor while facing the practical constraints of classroom assessment. Their reflections on how well the tools align with these \textit{dual} roles can inform better designs. Seeing this opportunity, we ask what these instructor–researchers can teach us about designing visualization tools for qualitative learning analytics. We pose three research questions, in order of increasing abstraction:

\phantomsection\label{rq1}\textbf{RQ1.} How do \edit{these instructor-researchers} evaluate the two platforms with respect to ease of use, usefulness, and \edit{alignment with their qualitative education research practices} when analyzing student assessment data?

\phantomsection\label{rq2}\textbf{RQ2.} How do these experts judge the appropriateness of frequency-based quantitative encoding for representing qualitative data?

\phantomsection\label{rq3}\textbf{RQ3.} How might these judgments inform the design of visualizations for qualitative educational data?

In this study we investigate these questions through a mixed-methods user study with STEM education researchers who also teach courses. We originally designed a comparative evaluation of two qualitative learning analytics platforms, recognizing that they serve distinct primary purposes. The comparison, however, surfaced an unanticipated tension: beneath the numerical usability ratings lay a genuine disagreement about whether frequency-based encoding can adequately represent qualitative data \cite{debatereview} . Taken together, this study offers the following contributions:

\vspace{-.5em}
\begin{itemize} \setlength\itemsep{0.1em}
    \item a mixed-methods evaluation, with STEM instructor--researchers, of two qualitative learning analytics platforms, using real-world student journals to assess ease of use, usefulness, and alignment with qualitative education research practices;
    
    \item an emergent characterization of epistemological stances among these participants regarding the use of frequency-based encodings for qualitative educational data; and

    \item design implications for visualization tools that seek to integrate quantitative techniques into the qualitative analysis of educational data.
\end{itemize}

\section{Related Work}
\label{sec:related_work}

\subsection{Bridging Visualization and Qualitative Research}
Visualization and qualitative approaches connect in several directions in the service of research, evaluation, and education ~\cite{losev2022embracing,diehl2020studying,Interplay,meng2026designing, diehl2021visguided}. For instance, Grounded theory, a qualitative research method, has been used within visualization research~\cite{Interplay, isenberg2008grounded} and while developing text-analytic interfaces ~\cite{chandrasegaran2017integrating}. 

Similar qualitative methods are often used to evaluate visualizations. Input from feedback sessions often takes written form~\cite{interview_study}. Text data is often collected as words directly through surveys, interviews, video transcripts, notes, or written feedback. Researchers code this text into themes that become concrete feedback and design implications, informing new tools and recommendations. However, qualitative reporting has tended to lean on quotes, tables, process models, and conceptual diagrams more than on data displays~\cite{verdinelli2013data, jaylo2026visualization}. These forms suggest an opportunity for innovation to support users in their qualitative analysis and reporting.

Quantitative ethnography is one of the most direct attempts to bridge this divide, developing statistical representations of coded discourse~\cite{shaffer2018epistemic, abramson2015beyond}. Its network-based methods illustrate that quantification can move beyond counts to focus on relationships. Epistemic network analysis (ENA) and ordered network analysis (ONA) quantify and visualize patterns of association among qualitative codes as weighted network models, with ONA additionally modeling the order in which those connections occur~\cite{tan2024epistemic}.

 In summary, qualitative approaches are increasingly found in visualization research and evaluation but reported with only a limited set of visual representations, and ongoing work with coded textual data further integrates these areas.

\subsection{Temporal Text Visualization}
We focus on text data with a time dimension, the case pertinent to qualitative learning analytics where an instructor tracks student writing over a semester. Foundational work established the streamgraph idiom for this task: ThemeRiver~\cite{havre2002themeriver} visualized thematic change across a document collection over time, and later systems such as TIARA~\cite{wei2010tiara} and TextFlow~\cite{cui2011textflow} extended it to track the evolution, birth, and merging of topics. WordStream~\cite{wordstream} refined this line of work by keeping the extracted keywords legible on screen, and the idiom has since been embedded in larger interactive dashboards~\cite{zhang2022will}, frequently for text analysis~\cite{EQSA, van2019hackernets, dang2020agasedviz}. Application examples span domain-specific tasks: CovidStream~\cite{baca2020covidstream} traces the evolution of emotions throughout the COVID-19 pandemic in Peru, and VisDmk~\cite{cao2022visdmk} represents the evolving sentiment and pop-up commentary of a video. The two platforms we evaluate, the Journal Data Dashboard~\cite{nguyen2021interactive} and WordStream Maker~\cite{nguyen2022wordstream}, put this idiom in the service of education. Despite WordStream's growing adoption, it has rarely been subjected to rigorous evaluation with domain experts who would actually use it for qualitative analysis, a gap our study aims to address.

\section{The Probes: Journal Data Dashboard and WordStream Maker}
\label{sec:background}

Qualitative learning analytics require visualizations that connect high-level trends with close reading, supporting both instructional feedback and in-depth research on student writing. WordStream~\cite{wordstream} provides a temporal overview by displaying changes in term prevalence. It combines a word cloud~\cite{viegas2009participatory} with the temporal layout of a streamgraph~\cite{harris1999information}, keeping extracted terms legible along a timeline. Section~\ref{subsection:wordstream} details the encoding.

We use two platforms built on this idiom as \textit{probes}~\cite{hutchinson2003technology} to elicit expert perspectives on visualization for qualitative learning analytics: the Journal Data Dashboard~\cite{nguyen2021interactive} for analyzing formative assessments and WordStream Maker~\cite{nguyen2022wordstream} for authoring custom visualizations. Both are open-source, browser-based applications. As summarized in Table~\ref{tab:design}, the probes differ in scope, data input, access to source text, and customization. Their distinct purposes make a direct feature-by-feature comparison \textit{inherently asymmetric}. We use these contrasts to examine what expert users value, trust, and critique at different stages of an educational data visualization workflow.

\subsection{Journal Data Dashboard}
\begin{figure}[h]
  \centering
  \includegraphics[width=\linewidth]{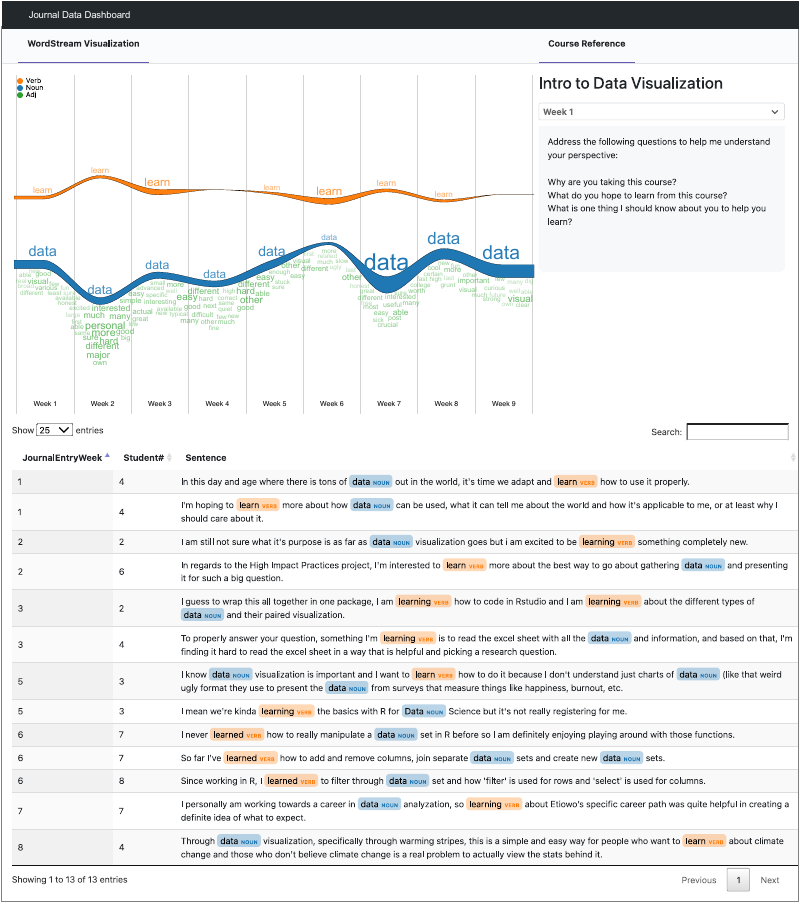}
  \vspace{-1.3em}
  \caption{Journal Data Dashboard~\cite{nguyen2021interactive} user interface for analyzing formative assessments}
  \label{fig:dashboard_background}
\end{figure}

Journal Data Dashboard~\cite{nguyen2021interactive}, hereafter \textbf{Dashboard}, is a multi-panel tool for exploring a fixed corpus of formative assessments. It links a WordStream to the weekly prompts that students answered and to their full responses. As shown Figure~\ref{fig:dashboard_background}, selecting a word retrieves the responses that contain it. These links support \textit{insight extraction} by allowing users to move from aggregate patterns to individual student writing.

\vspace{-.1em}
\vspace{0em}

The Dashboard is available as a web-based application at
\href{https://huyen-nguyen.github.io/journal-dashboard/wordstream}{\texttt{huyen-nguyen.github.io/journal-dashboard/wordstream}}, with an accompanying tutorial\footnote{\href{https://huyen-nguyen.github.io/journal-dashboard/tutorials}{\texttt{huyen-nguyen.github.io/journal-dashboard/tutorials}}} and a demonstration video.\footnote{\href{https://huyen-nguyen.github.io/journal-dashboard/video}{\texttt{huyen-nguyen.github.io/journal-dashboard/video}}}

\subsection{WordStream Maker}
\begin{figure}[h]
  \centering
  \includegraphics[width=\linewidth]{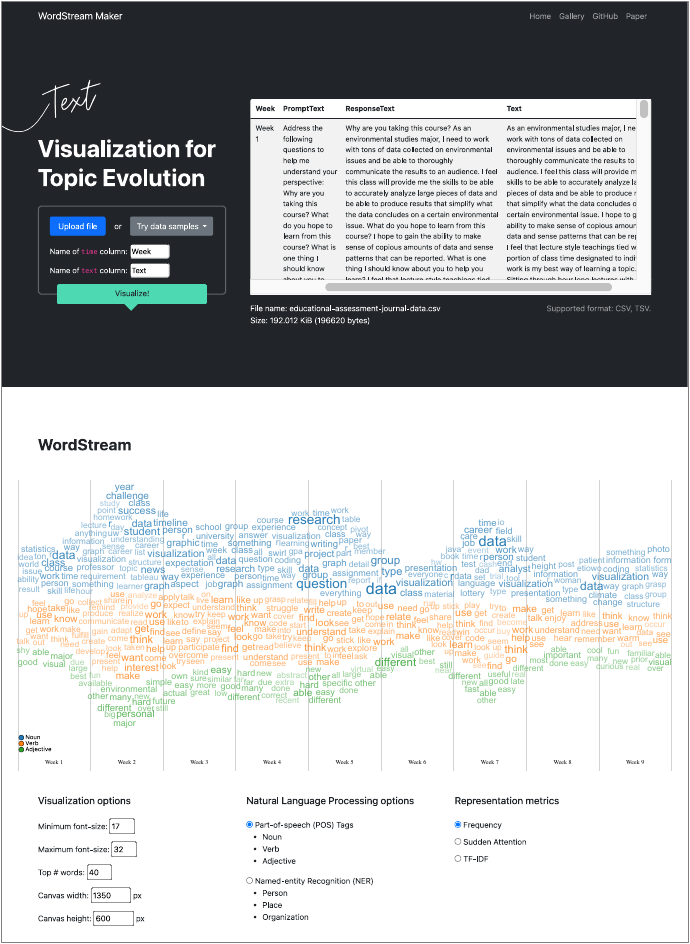}
  \caption{WordStream Maker~\cite{nguyen2022wordstream}, an end-to-end tool for authoring custom visualizations.}
  \label{fig:maker_background}
\end{figure}

WordStream Maker~\cite{nguyen2022wordstream}, hereafter \textbf{Maker}, is a general-purpose, no-code tool for \textit{visualization creation}, as shown in Panel B of Figure~\ref{fig:maker_background}. Its graphical interface accepts user-provided tabular text data and combines data wrangling, natural language processing (NLP), and WordStream rendering in a single pipeline. Users can select part-of-speech categories and word-scoring methods. The resulting aggregate views, however, do not provide linked access to the source responses.

Maker includes built-in examples and is available online at
\href{https://huyen-nguyen.github.io/maker}{\texttt{huyen-nguyen.github.io/maker}}.
In this study, participants used the ``Education Assessment'' sample, which contains the same student reflection corpus presented in the Dashboard.

{
\renewcommand{\arraystretch}{1.4}
\begin{table}[h]
  \caption{Key design contrasts between the two probes, Dashboard~\cite{nguyen2021interactive} and Maker~\cite{nguyen2022wordstream}. Each tool supports a different activity within an analytical workflow.}
  \label{tab:design}
  \centering
  \footnotesize
  \begin{tabularx}{\linewidth}{p{1.4cm} X X}
    \toprule
     & \textbf{Dashboard~\cite{nguyen2021interactive}}
     & \textbf{Maker~\cite{nguyen2022wordstream}} \\
    \midrule
    Scope         & Domain-specific
                  & General-purpose \\
    Data          & Fixed assessment corpus
                  & User-uploaded at runtime \\
    Linked text   & Verbatim responses
                  & None \\
    Primary role  & Insight extraction
                  & Visualization creation \\
    Customization & Limited, preset
                  & User-selectable parameters \\
    \bottomrule
  \end{tabularx}
\end{table}
}

\subsection{Shared Visualization Idiom: WordStream}
\label{subsection:wordstream}
Within the a WordStream~\cite{wordstream} view, which are embedded in both Figures~\ref{fig:dashboard_background} and~\ref{fig:maker_background}, words and phrases are sized by a computed score at each time point. Both probes use word frequency by default; Maker also offers sudden attention and TF--IDF as alternative scoring methods. The appropriateness of frequency-based encoding for qualitative educational data is the focus of this research study.

The shared idiom provides a common basis for exploring temporal patterns, while the platforms offer different ways to interpret those patterns. In particular, keeping extracted terms visible in WordStream differs from providing access to their full context: the Dashboard supplies that context through linked responses, whereas Maker presents an aggregate view.

\section{Methods}
\label{sec:methods}

\subsection{Conceptual Framing}
This work is guided by the Technology Acceptance Model (TAM)~\cite{davis1989perceived}, which suggests that users are more likely to adopt a technology based on its perceived ease of use and perceived usefuless. Alongside TAM, we incorporate the concept of \textit{context of use}~\cite{isenberg2008grounded, tory2014user} to better understand the circumstances surrounding the use or potential use of such tools. Specifically, we frame the visualization platforms as technology probes~\cite{hutchinson2003technology} through which participants can reflect on existing workflows and possible changes to their practice. Together, these perspectives guide our use of both platforms to elicit experts' perception on their usefulness and ease of use, in relation to their current work contexts and potentially explore how the tools might support their practices.

\begin{figure*}[!t]
  \centering
  \begin{tikzpicture}[
  font=\sffamily\large,
  >={Stealth[length=2.6mm]},
  base/.style={rounded corners=2pt, draw=PalBlack!40, line width=.6pt,
               align=center, inner sep=5pt},
  head/.style={base, font=\sffamily\bfseries\large, text=white, inner sep=5pt},
  setup/.style={base, fill=SetupLightBg, draw=SetupBorder,
                text width=36mm, font=\sffamily\large},
  setuphead/.style={head, fill=SetupNavy, draw=SetupNavy, text width=36mm},
  qbox/.style ={base, fill=QualFill,    draw=QualStroke,
                text width=26mm, font=\sffamily\large},
  theme/.style={base, fill=QualFill,      draw=QualStroke,
                text width=30mm, font=\sffamily\large},
  tbox/.style ={base, fill=QuantFill,    draw=QuantStroke,
                text width=26mm, font=\sffamily\large},
  pool/.style ={base, fill=AnalysisFill, draw=AnalysisStroke,
                text width=34mm, font=\sffamily\large},
  integ/.style={base, fill=AnalysisFill, draw=AnalysisStroke,
                text width=34mm, font=\sffamily\large},
  imp/.style={base, fill=AnalysisFill, draw=AnalysisStroke,
                text width=34mm, font=\sffamily\large},
  integhead/.style={head, fill=IntegHead, draw=IntegHead, text width=34mm},
  rq/.style={base, rounded corners=7pt, fill=RqFill, draw=RqStroke,
             font=\sffamily\bfseries\large, text width=20mm, inner sep=3pt},
  flow/.style={->, line width=.8pt, draw=PalBlack!45},
  qflow/.style={->, line width=.8pt, draw=QualStroke},
  tflow/.style={->, line width=.8pt, draw=QuantStroke},
  rqarrow/.style={->, line width=.6pt, draw=RqStroke},
  lanelab/.style={font=\sffamily\bfseries\large, align=center},
  clabel/.style={font=\sffamily\bfseries\large, text=PalBlack!60},
  legend/.style={draw=none, fill=none, align=left, font=\sffamily\large},
]

\def\ty{-1.7}       %
\def\qy{-4.1}    %
\def\qylabel{-4.5}    %

\node[setuphead] (suh) at (0,.5) {Setup \& instrument};
\node[setup, below=4mm of suh] (corpus)
  {\textbf{Underlying dataset} \\ Students \mbox{written} reflections \\ (Data Vis course)};
\node[setup, below=4mm of corpus] (instr)
  {\textbf{Qualtrics \mbox{survey}} \\comparing \Dashboard{} \& \Maker{}  \\
  ${\approx}1$\,hr.
  };
\node[setup, below=4mm of instr] (sample)
  {\textbf{Participants:} STEM \emph{instructor--researchers};\\ $n{=}10$};
  
\draw[flow] (suh)--(corpus); \draw[flow] (corpus)--(instr); \draw[flow] (instr)--(sample);

\node[tbox](c1d) at (6.3,\ty)
  {\textbf{Likert} \\ \hyperref[g:likert]{L1--L6} \\[1mm] for \Dashboard{}};
\node[tbox, right=8mm of c1d] (c1m)
  {\textbf{Likert} \\ \hyperref[g:likert]{L1--L6}\\[1mm] for \Maker{} \;};

\node[tbox, right=22mm of c1m] (c2d)
  {\textbf{Likert} \\ \hyperref[g:likert]{L1--L6}\\[1mm] for \Dashboard{}};
\node[tbox, right=8mm of c2d] (c2m)
  {\textbf{Likert} \\ \hyperref[g:likert]{L1--L6}\\[1mm] for \Maker{}};

\draw[tflow] (c1d)--(c1m);
\draw[tflow] (c1m)--(c2d);
\draw[tflow] (c2d)--(c2m);

\node[qbox]  (c1q) at ($(c1d)!(0,\qy)!(c1d)$)  {\textbf{Open questions} \\ \hyperref[g:open]{O1--O6}};
\node[theme] (c1t) at ($(c1m)!(0,\qy)!(c1m)$) {\textbf{Preliminary \mbox{thematic} \\ analysis}};
\node[qbox]  (c2q) at ($(c2d)!(0,\qy)!(c2d)$) {
\textbf{Open} \\ \hyperref[g:open]{O1--O6} \& \\[1 mm] \textbf{Confirmatory questions} \\  \hyperref[g:confirm]{C1--C5}};

\node[theme] (c2t)
  at ($(c2m)!(0,\qy)!(c2m)$)
  {\textbf{Final \mbox{thematic} analysis} w/ \\Member check};
\draw[qflow] (c1q)--(c1t);
\draw[qflow] (c1t)--(c2q)
  node[midway, above=1mm,
       font=\sffamily\large, text=QualStroke, align=center]
  {preliminary\\themes};
\draw[qflow] (c2q)--(c2t);

\node[integhead] (inh) at (24,.3) {Integration};

\node[pool, below=4mm of inh, minimum height=14mm] (pk)
  {\textbf{Comparative analysis}\\ pooled $n{=}10$:\\
   \textbf{Dashboard} vs.\ \textbf{Maker}};

\node[integ, below=5mm of pk, minimum height=17mm] (interp)
  {\textbf{Mixed-methods} \textbf{interpretation}\\ ratings  $+$ themes\\};
\node[integ, below=4mm of interp, minimum height=11mm] (imp)
  {\textbf{Design \mbox{implications}}};

\draw[flow] (inh)--(pk);
\draw[flow] (pk)--(interp);
\draw[flow] (interp)--(imp);

\draw[tflow] (c2m.east) to[out=5, in=180, looseness=.4] (pk.west);

\draw[qflow] (c2t.east) to[out=5,in=180, looseness=1] (interp.west);

\node[rq, below=15mm of c2t] (rq2) {
\textbf{\hyperref[rq1]{RQ1}}\,\& 
\textbf{\hyperref[rq2]{RQ2}}};
\draw[rqarrow] (c2t.south) -- (rq2.north);

\node[rq, below=10mm of imp] (rq3) {\hyperref[rq3]{RQ3}};

\draw[rqarrow] (imp.south) -- (rq3.north);

\node[rq] (rq1) at (25, 2) {\hyperref[rq1]{RQ1}};
\draw[rqarrow] (pk.east) to[out=0,in=0, looseness=.8] (rq1.east);

\node[lanelab, text=QuantStroke, anchor=east, align = left] at (4.45,\ty) {Quantitative\\ strand};
\node[lanelab, text=QualStroke,  anchor=east, align =left] at (4.25,\qylabel ) {Qualitative\\ strand};

\begin{scope}[on background layer]
  \node[draw=SetupBorder, dashed, rounded corners=5pt, line width=.8pt,
        fill=SetupZoneBg, fit=(suh)(corpus)(instr)(sample), inner sep=7pt] (sb) {};
  \node[draw=PalBlack!40, dashed, rounded corners=5pt, line width=.8pt, fill=PalBlack!2,
        fit=(c1q)(c1t)(c1d)(c1m), inner sep=8pt] (cb1) {};
  \node[draw=PalBlack!40, dashed, rounded corners=5pt, line width=.8pt, fill=PalBlack!2,
        fit=(c2q)(c2t)(c2d)(c2m), inner sep=8pt] (cb2) {};
  \node[draw=AnalysisStroke, dashed, rounded corners=5pt, line width=.8pt,
        fill=AnalysisZoneBg, fit=(inh)(interp)(pk)(imp), inner sep=8pt] (ib) {};
\end{scope}

\node[clabel, anchor=south] at (sb.north)  {\vphantom{Ag}Study design};
\node[clabel, anchor=south] at (cb1.north) {\vphantom{Ag}Cycle 1\ (first $n{=}4$ participants)};
\node[clabel, anchor=south] at (cb2.north) {\vphantom{Ag}Cycle 2\ ($n{=}6$ new participants; 10 total)};
\node[clabel, anchor=south] at (ib.north)  {\vphantom{Ag}Analysis \&\ interpretation};

\draw[tflow] (sb.east) to[out=0,in=180]  (c1d.west);
\draw[qflow] (sb.east) to[out=-15,in=180] (c1q.west);

\node[legend, anchor=north west] at ([yshift=-5mm]sb.south west) {%
  \textbf{Mixed-methods design.}\quad
  \textbf{\hyperref[rq1]{RQ1}}\,:\,Ease of use, usefulness, fit to practice;\quad
  \textbf{\hyperref[rq2]{RQ2}}\,:\,Appropriateness of frequency-based encoding;\quad
  \textbf{\hyperref[rq3]{RQ3}}\,:\,Design implications.};
\end{tikzpicture} %
  \caption{Study procedure with a two-phase mixed-methods design with ten STEM instructor-researchers. All participants worked on the same student reflection corpus within a survey, which included a quantitative strand (Likert ratings \hyperref[g:likert]{L1--7} for Dashboard \& Maker tool) and a qualitative strand (open-ended responses \hyperref[g:open]{O1--O6}, plus confirmatory items \hyperref[g:confirm]{C1--C5} in Cycle 2). Themes from Cycle 1 ($n{=}4$) were checked and refined in Cycle 2 ($n{=}6$), producing four themes (T1–T4) that were interpreted together with tool comparisons ($n{=}10$) to answer \hyperref[rq1]{RQ1}, \hyperref[rq2]{RQ2}, and \hyperref[rq3]{RQ3}.
}
  \label{fig:procedure}
\end{figure*}

\subsection{Study Design}

A mixed-methods design  was used following an initial pilot study ~\cite{creswell2017designing, nguyen2021interactive}. We included two qualitative-quantitative cycles with confirmatory checking. This design offers flexibility and supports emergent findings, appropriate for the research pursuit. The study procedure overview is shown in Figure~\ref{fig:procedure}.

A quantitative component compared the two platforms, Dashboard and Maker, on perceived ease of use and usefulness. Two platforms serve distinct primary purposes, so we positioned them not as direct competitors but as probes into two stages of an analytical workflow: creating a visualization (Maker) from raw data and extracting insight from an interactive multi-view with linked text excerpts (Dashboard). The quantitative approach primarily addresses \hyperref[rq1]{RQ1}, providing an evaluation by expert users.

A concurrent qualitative component used thematic analysis to explore participants’ experiences and contexts of use beyond what an agreement scale can capture. A qualitative approach addresses \hyperref[rq1]{RQ1} and \hyperref[rq2]{RQ2}, providing judgment of appropriateness.

The quantitative and qualitative findings were integrated in a final interpretation research stage. This stage supports evidence integration, providing the basis of design recommendations, addressing \hyperref[rq3]{RQ3}.

\subsection{Participants}

We recruited ten STEM education researchers with expertise in both qualitative research methods and classroom assessment. These instructor-researchers were sampled purposefully for their dual expertise. We refer to participants as Experts~1--10; this numbering reflects our internal labeling and is not ordered meaningfully. The first analysis cycle included four participants (Experts~1, 7, 8, and~9), and the second cycle comprised all ten participants, with six newly recruited (Experts~2, 3, 4, 5, 6, and~10). Participants received a \$100 (USD) gift card for their time.

\subsection{Tasks and Apparatus}
\label{sec:tasks}
Each participant used both platforms, performed the assigned tasks, and completed a Qualtrics survey in approximately one hour. During the task, participants first explored the Dashboard and then explored the Maker, both of which accessed the same underlying ``Education Assessment'' sample dataset, consisting of 62 weekly reflections written by students in a data visualization course. Both the task tutorial and the complete instruments are provided in the Supplementary Material.

The survey questions were grounded in the TAM and context of use, measuring perceived usefulness and easy to use ~\cite{davis1989perceived}. 
The survey included seven rating statements for each platform and six open-response questions, and for six participants in the second cycle, five additional confirmatory questions were included.
All participants rated each platform on the same seven agreement statements (five-point Likert scale: \emph{strongly disagree}, \emph{disagree}, \emph{neither agree nor disagree}, \emph{agree}, \emph{strongly agree}). The seven Likert-rating statements for evaluating both platforms were as follows:

\phantomsection\label{g:likert}
\begin{enumerate}\itemsep0em
\item[\textbf{L1}]\phantomsection\label{q:L1}\textbf{Details.} The app allows me to explore the details of the students' responses.
  \item[\textbf{L2}]\phantomsection\label{q:L2}\textbf{Trends.} The app allows me to identify broad trends among the students' responses.
  \item[\textbf{L3}]\phantomsection\label{q:L3}\textbf{Progress.} The app helps me draw conclusions related to student progress.
  \item[\textbf{L4}]\phantomsection\label{q:L4}\textbf{Ease of use.} It is easy to use the app.
  \item[\textbf{L5}]\phantomsection\label{q:L5}\textbf{Exploration.} The app makes it easier to explore student ideas.
  \item[\textbf{L6}]\phantomsection\label{q:L6}\textbf{Development.} The app gives me insight into the development of student ideas over time.
  \item[\textbf{L7}]\phantomsection\label{q:L7}\textbf{Diversity.} The app lets me see the diversity in students' responses and experiences.
\end{enumerate}

\phantomsection\label{g:open}
\noindent The six open-response questions were as follows:
\begin{enumerate}\itemsep1.2pt
  \item[\textbf{O1}]\phantomsection\label{q:O1}\textbf{Preference.} Which of the two tools did you prefer? Why?
  \item[\textbf{O2}]\phantomsection\label{q:O2}\textbf{Trust.} Compare the data presented by both tools. To what extent do you trust the visualization(s) provided? Explain.
  \item[\textbf{O3}]\phantomsection\label{q:O3}\textbf{Concerns.} What concerns do you have about how the visualizations represent the data? Explain.
  \item[\textbf{O4}]\phantomsection\label{q:O4}\textbf{Workflow change.} Thinking about how you usually analyze and present qualitative data, how would these tools change the way you work, if at all?
  \item[\textbf{O5}]\phantomsection\label{q:O5}\textbf{Desired features.} What features (current or additional) would make you more likely to use these tools? Explain.
  \item[\textbf{O6}]\phantomsection\label{q:O6}\textbf{Other thoughts.} Please share any other thoughts you have about the tools.
\end{enumerate}

\noindent The seven rating statements (\hyperref[g:likert]{L1--7}) and the preference, trust, and workflow change questions (\qref{O1},\qref{O2},\qref{O4}) target \hyperref[rq1]{RQ1} to understand the participants' evaluation of the tools; the concerns and desired features questions (\qref{O3},\qref{O5}) elicit their judgments about frequency-based encoding central to \hyperref[rq2]{RQ2}, relating the platforms to the representational format of qualitative data. In parallel, the workflow change and the desired features questions (\qref{O4},\qref{O5}) motivate \hyperref[rq3]{RQ3}.

To test the preliminary themes emerging from Cycle~1, the six Cycle~2 participants additionally responded to five \textit{confirmatory} statements, each rated on the same five-point scale and accompanied by a free-text explanation. The five confirmation themes were as follows:

\phantomsection\label{g:confirm}
\begin{enumerate}\itemsep1pt
  \item[\textbf{C1}]\phantomsection\label{q:C1}\textbf{Rich context.} I valued functions that helped me explore the rich context of relationships in the qualitative data.
  \item[\textbf{C2}]\phantomsection\label{q:C2}\textbf{Compare in context.} I valued functions for comparing and viewing data in context.
  \item[\textbf{C3}]\phantomsection\label{q:C3}\textbf{More control.} More functions would allow me to increase my control to analyze data in context.
  \item[\textbf{C4}]\phantomsection\label{q:C4}\textbf{Distrust under uncertainty.} I distrusted the tool when the functions evoked uncertainty.
  \item[\textbf{C5}]\phantomsection\label{q:C5}\textbf{Specific contexts.} I considered specific contexts (e.g., teaching, assessment, research) when I evaluated the tools.
\end{enumerate}

These confirmatory items functioned as a member-checking step, returning the candidate themes underlying \hyperref[rq1]{RQ1} and \hyperref[rq2]{RQ2} to participants for agreement and explanation. The complete instruments are provided in the Supplementary Material.

\subsection{Quantitative Analysis}

Likert responses were summarized per statement (\hyperref[g:likert]{L1--L7}) and per tool (Dashboard vs. Maker) for visual analysis (Figure~\ref{fig:per-item}, Table~\ref{tab:items}); the small purposeful sample ($n{=}10$) precludes inferential statistics~\cite{chen2021methods}. We report group-level means per statement and per tool, treating the ratings as ordinal indicators of perceived ease of use and perceived usefulness rather than as a basis for hypothesis testing. This component provides the comparative evaluation that answers \hyperref[rq1]{RQ1}.

\subsection{Thematic Analysis}
A thematic analysis was selected to conduct the concurrent qualitative study due to its ability to explore insightful patterns of sentiments and experiences not captured on an agreement scale. We performed a thematic analysis of written responses to the open-format questions following a six-step approach of coding data to develop themes iteratively over two cycles ~\cite{braun2006using,braun2023toward} with the aid of Dedoose~\cite{dedoose2021dedoose} (see Supplementary Material for details). To ensure the credibility of the interpretation, the last author (CMT) and a research student met weekly to discuss the coding process, emerging themes, and the meanings of the definitions. Before proceeding with preliminary findings, the first author (HNN) and second author (KAB) triangulated the analysis by verifying, modifying, or denying the defined themes with specific examples from the experts' writing.

After the second cycle, the six phases of thematic analysis were conducted again with the new participant data. During this iteration, three new team members, including the fourth author (KAT), were involved before seeking consensus with the full group. The second cycle led to the removal of one preliminary theme from consideration and the addition of a new theme, Theme~4, based on ample evidence (detailed in the Supplementary Material). The full team then debriefed about the themes and agreement-data visuals produced, providing additional insights, clarification, and limitations for the final themes, and integrating the two data sets for interpretation. As with other qualitative approaches, we did not intend to make generalizable claims; instead, we aimed to produce highly contextualized, generative knowledge from our sample in order to further the design and provide deep insights for future studies~\cite{guba2005paradigmatic}.

\section{Results}
\label{sec:result}

\subsection{Quantitative Findings on Perceived Ease of Use and Usefulness (\hyperref[rq1]{RQ1})}
\label{subsec:quant_result}

\begin{figure*}[!t]
  \centering
  \includegraphics[]{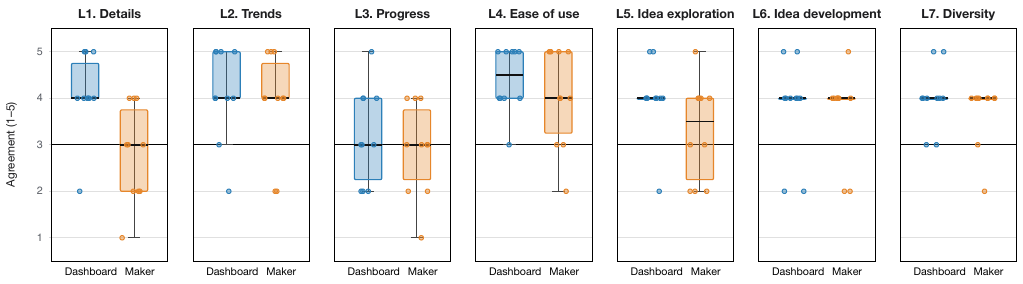}
\caption{Agreement ratings (1--5, 5 = strongly agree) for the Dashboard (blue) and Maker (orange) on seven \hyperref[g:likert]{Likert items (L1--L7)}, across all ten participants. Boxes show the median and interquartile range; points represent individual participants with horizontal jitter to reduce overplotting. Given the small sample (n = 10), individual data points are shown on each box plot, and quartiles are descriptive rather than inferential.}
\label{fig:per-item}
\end{figure*}

Across all ten participants, the Dashboard was rated at or above the Maker on every Likert item. Table~\ref{tab:items} reports per-item means derived directly from the study data, and Figure~\ref{fig:per-item} shows detailed agreement ratings on each of the seven Likert items. Given the small sample $n{=}10$, we overlay each participant's individual rating (jittered to prevent overplotting) on the box plots, keeping the per-participant distribution as the primary basis for interpretation.

\begin{table}[htbp]
  \centering
  \caption{Per-item mean ratings (1--5 scale) for the Dashboard and the
           Maker across all $n{=}10$ participants. The mean of Dashboard
           is at or above the Maker's on every item.}
  \label{tab:items}
  \begin{tabular}{llccc}
    \toprule
    Item & Description & Dashboard & Maker & $\Delta$ \\
    \midrule
    \qref{L1} & Details     & 4.10 & 2.80 & $+$1.30 \\
    \qref{L2} & Trends      & 4.10 & 3.90 & $+$0.20 \\
    \qref{L3} & Progress    & 3.20 & 2.90 & $+$0.30 \\
    \qref{L4} & Ease of use & 4.40 & 4.00 & $+$0.40 \\
    \qref{L5} & Idea exploration & 4.00 & 3.30 & $+$0.70 \\
    \qref{L6} & Idea development & 3.80 & 3.70 & $+$0.10 \\
    \qref{L7} & Diversity   & 4.00 & 3.70 & $+$0.30 \\
    \midrule
    ---        & \textbf{Overall mean} & \textbf{3.94} & \textbf{3.47} & $+$\textbf{0.47} \\
    \bottomrule
  \end{tabular}
\end{table}

\begin{figure}[!t]
  \centering
  \includegraphics[width=.858\linewidth]{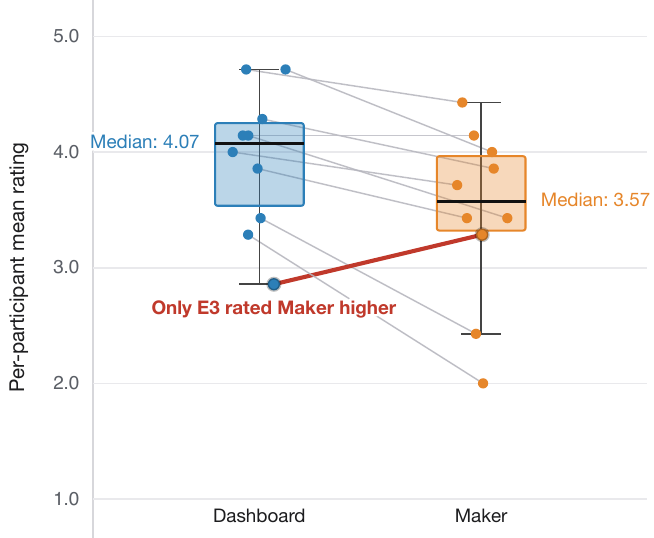}
\caption{Per-participant mean agreement ratings (\textbf{averaged across the 7 Likert items}) for the Dashboard (blue) and Maker (orange). Each gray line represents a participant, linking their paired scores. Most lines slope downward, showing a general preference for the Dashboard; the red line marks the only exception (E3). Similar to Figure~\ref{fig:per-item}, with n = 10 the boxplots only show descriptive medians rather than inferential.}
\vspace{-1em}
\label{fig:per-user}
\end{figure}

The overall mean was 3.94 for the Dashboard and 3.47 for the Maker; expressed as a share of \emph{agree} or \emph{strongly agree} ratings, 77\% of Dashboard ratings (54/70) were positive versus 60\% for the Maker (42/70). The largest gap is on exploring details (\qref{L1}, $\Delta = 1.30$) and almost no difference on broad trends (\qref{L2}, $\Delta = 0.20$) and development over time (\qref{L6}, $\Delta = 0.10$). Both platforms received their highest ratings on ease of use (\qref{L4}: Dashboard 4.40, Maker 4.00) and their lowest on student progress (\qref{L3}: Dashboard 3.20, Maker 2.90). Variance across participants was substantial on several items, reflecting the diversity of disciplinary stances discussed in the following sections.

At the per-participant level (Figure~\ref{fig:per-user}), the same pattern holds when ratings are collapsed across items: nine of the ten participants gave the Dashboard a mean agreement rating at or above their Maker rating, yielding the predominantly downward-sloping connecting lines, with Expert~3 the sole exception (red line). The descriptive medians of the per-participant means are 4.07 for the Dashboard and 3.57 for the Maker, and the Maker's distribution extends further toward the lower end of the scale, where two participants sit near or below the midpoint.

Given the purposeful but small sample, we read these as descriptive indicators rather than evidence of a reliable effect. In the following section, tool preference (\qref{O1}) reflects the ratings: seven participants preferred the Dashboard, two preferred the Maker (Experts~3 and~4, both citing its side-by-side, simultaneous temporal display), and one (Expert~8) was mixed, valuing the Maker for quantitative trend-spotting while preferring the Dashboard's in-context word search.

\subsection{Themes 1--3: Tool Experience and Disciplinary Context of Use (\hyperref[rq1]{RQ1})}
\label{subsec:themes}

Three themes characterize how participants experienced the tools and situated them in their own practice. The five confirmatory items (\hyperref[g:confirm]{C1--C5}), answered by the six Cycle~2 participants, served as a member check on these themes; agreement was strong for the items underlying Themes~1--3 (\hyperref[g:confirm]{C1,C2,C3,C5}). We report Theme 4 in a dedicated Section~\ref{subsec:dissensus} to answer \hyperref[rq2]{RQ2}.

\theme{Theme 1: Enhancing navigation and exploration of qualitative patterns across scales of context.} Users valued features that enhanced ease of use by making it easier to compare and examine qualitative data in context at different scales of analysis. Specifically, they appreciated being able to search responses for words and phrases through filters, view trends and patterns through visualizations, and, for some, associate those words and phrases with quantitative trends. Expert~1 wrote: 
\begin{quote}
\inquote{I preferred [Dashboard] because it provided both a traditional wordcloud in addition to the weekly visualization. I also like the feature of being able to \textbf{select words in combination} with one another to view the responses where they were used together.} (Cycle~1, \qref{O1})
\end{quote}
Expert~7  similarly valued the navigability from trend to text: 
\begin{quote}
    \inquote{The [Dashboard] is easier to navigate, I can easily see the \textbf{pattern of words per questions}, I can navigate from the summary to the wordstream and see changes in trends.}
    (Cycle~1, \qref{O1})
\end{quote}

Their feedback focused on the specific functionality, interactions, and visual elements of each platform's existing design. The member check reinforced this; explaining strong agreement with \qref{C2}, Expert~6 (Cycle~2) noted, \inquote{``Context gives a great deal of meaning to these topics/phrases.''} Across responses, a common thread was the desire for autonomy where participants wanted to explore the dataset on their own terms using a variety of representations. 

This autonomy extended to the unit of analysis itself. Participants wanted to recompose the same corpus around units other than the whole class. Expert~1 wanted \inquote{``to select the responses of only one student over time''} (Cycle~1, \qref{O5}); Expert~3 wanted students' responses \inquote{``grouped by question''} (Cycle~2, \qref{O3}); Expert~2 wanted \inquote{``a way to look at how many students are saying particular phrases (not just how common the words are overall)''} (Cycle~2, \qref{O5}); and Expert~10 wanted to track \inquote{``multiple words within the same category''} together (Cycle~2, \qref{O5}).

Ultimately, the perceived ease of searching, comparing, organizing, and otherwise manipulating the visualization to make sense of patterns was the primary basis on which users valued the WordStream platforms' features. 

\theme{Theme 2: Extending analytic workflows through user control.}
Users evaluated the tools' usefulness largely through the features they wanted added that would extend the visualizations into a fuller analytic pipeline, from data ingestion and pre-processing through coding, quantification, and integration with established software.
Several wanted the tools to ingest their own data. Expert~10  wrote:

\begin{quote}
\inquote{The functionality where I can \textbf{upload my own data} into this dashboard interface would be the icing on the cake for rapidly understanding my data in the early stages.}
(Cycle~2, \qref{O1})
\end{quote} 

Others asked for preprocessing capabilities that would clean inputs before visualization; Expert~1 (Cycle~1, \qref{O5}) wanted \inquote{``the ability to filter out some words,''} noting that prompt words tend to repeat across student responses and obscure the substantive vocabulary worth examining.

A second strand of suggestions pushed the tools downstream, toward existing analytic practice. Expert~2 saw the visualizations as a starting point for qualitative coding:
\begin{quote}
\inquote{These tools would be an excellent way to quickly look for common codes for further analysis... helpful for open coding and for looking for trends in how particular topics are discussed over time.} 
(Cycle~2, \qref{O4})
\end{quote} 

Expert~5 asked for direct interoperability with established software: 
\begin{quote}
    \inquote{I'd almost want it to merge with a tool like SPSS text analysis --- so analyze the text, tag it, and display the frequency of those tags.} (Cycle~2, \qref{O5})
\end{quote}

Expert~10 likewise wanted quantitative readouts attached to the qualitative view, suggesting instance counts for each noun, verb, and adjective; and Expert~1 called for the platforms themselves to be consolidated, asking that \inquote{``the customization of Maker [be] embedded within Journal [Dashboard].''}

The two tools' strengths were complementary: the Dashboard led precisely on the in-context items the Maker lacks (details \qref{L1}, exploration \qref{L5}), while the Maker offered the customization the Dashboard lacks. Participants asked for the combination, as in Expert~1's request to consolidate the platforms and Expert~10's request to upload their own data into the Dashboard.

Across these suggestions, users considered the tools as candidate components of their existing analytic practice rather than as a standalone artifacts. Usefulness, in this case, was a function of how readily the visualizations could be folded into the ingestion, preprocessing, coding, quantification, and reporting routines users already had. 

In the member check, participants framed this desire as one of more control. This was evident in ratings of \qref{C3}, which drew agreement or strong agreement from five of six Cycle~2 participants. From written responses, `more control' was specified as control over the analytic workflow surrounding the tools rather than the exploration within the visualization that was noted in Theme 1.

\theme{Theme 3: Evaluating features through a deep context-of-use perspective across disciplinary domains.} Participants evaluated the tools within their specific disciplinary activities: teaching, assessment, and research, demonstrated through their comments closely linked to these disciplinary uses. They heavily reference their practices in examining classroom assessments or conducting qualitative research when evaluating the tools, as shown in an excerpt from Expert~6:
\begin{quote}
    \inquote{There is certainly a need for tools like these. However, the demands placed on a tool meant for qualitative data analysis are complex and vary greatly from project to project. I'm inclined to think increasing user customization is going to be key for tools like this.}
    (Cycle~2, \qref{O6})
\end{quote}
This comment reflects familiarity with qualitative research practice and recognition that the highly variable demands placed on technology to accommodate diverse methods pose major limitations for any visualization tool. 

Our participants shared the strengths and weaknesses of the platforms related to their various roles and uses. For example, when asked to comment on an initial draft of this theme, Expert 5  made the \textbf{dual role} of educator and researcher explicit: 
\begin{quote}
\inquote{I always have this in mind - would this help me in teaching (e.g., give me fast feedback on my students' thinking), in research ([...] looking at students' self-reported study skills - this type of tool could be really helpful here, giving us some initial ideas of study approaches students are using).}
(Cycle~2, \qref{C5})
\end{quote}

Envisioned research uses clustered around early-stage analysis; Expert~4 (Cycle~2, \qref{O4}) saw a fit for \inquote{``a great addition to initial qualitative coding to get an idea of what responses contain. Could be useful in building a codebook.''} Confirmatory item \qref{C5} drew agreement or strong agreement from all six Cycle~2 participants, each of whom named teaching, assessment, or research as the lens through which they judged the tools.

\subsection{Theme T4: An Epistemological Dissensus over Frequency-Based Encoding (\hyperref[rq2]{RQ2})}
\label{subsec:dissensus}

This Section is dedicated to \textbf{Theme T4}, which emerged only in Cycle~2 and prompted the addition of \hyperref[rq2]{RQ2}. \textbf{Theme T4} was a genuine disagreement about whether a frequency-based encoding can carry qualitative meaning. Participants ranged from skepticism to enthusiasm, with several occupying a pragmatic stance in the middle. At the more skeptical end, Expert~8 elaborated on shifting toward purpose-dependent uses:
\begin{quote}
    \inquote{Word frequency is not a qualitative research tool, but can direct researchers to transcripts or videos to help identify patterns.}
    (Cycle~1, \qref{O3})
\end{quote}
\noindent They later add: 
\begin{quote}
\inquote{I'm struggling to see how the frequency of a word can be used for qualitative data analysis where the researcher wants to know how or why, or even a common theme.} 
(\qref{O6})
\end{quote}
 
Expert~5 (Cycle~2, \qref{O3}) made the loss-of-context argument concretely: \inquote{``Word count/frequency doesn't actually tell us a lot. Words in context have a lot more meaning.''} Expert~2 named the long-tail risk: 

 \begin{quote}
      \inquote{While I think the tool would be very helpful in looking for general trends, sometimes \textbf{the most important \mbox{factors} are those that are not as common.}
      }
 (Cycle~2, \qref{O3})
 \end{quote} 

Expressing a similar concern on rare but critical responses, \mbox{Expert~3}  wrote: 

 \begin{quote}
 \inquote{I also worry about losing critical responses that are less frequent [...] I might expect to see a diversity of responses, and I would wonder about the bias introduced in this methodology.}
 (Cycle~2, \qref{O4})
  \end{quote} 
Alongside the Diversity item (\qref{L7}), both tools scored only moderately, which shows a limitation in how they represent data and affects how fairly we can analyze student voices.

A pragmatic middle treated frequency as helpful but partial. Expert~7 (Cycle~1, \qref{O2}) observed: \inquote{``Frequency of words can be helpful, but in my experience, short phrases can be more informative to capture students' ideas.''} Expert~10 located the tension in disciplinary reception rather than in the encoding itself:
\begin{quote}
\inquote{I would love for us to be operating in a world where visualizations in combination with student response data is sufficient for our colleagues to accept the significance of our findings, however, many still want a number.} (Cycle~2, \qref{O3})
\end{quote}

Even skeptical viewpoints found value in frequency as a \emph{locator} that routes attention back to close reading. Expert~8, quoted above, saw frequency as a way to direct researchers to the underlying transcripts, and described it as \inquote{``locator tools to direct me to where I will apply other methodologies''} (Cycle~1, \qref{O4}). Experts~2 and~4 likewise saw a role in early open coding and codebook building (Section~\ref{subsec:themes}).

At the enthusiastic end, Expert~9 stated: 
\begin{quote}
\inquote{I think you all have just revolutionized the future of qualitative coding, thematic analysis, and just how qualitative data is viewed and shared in general! I wish I had access to this tool when I was a graduate student!} 
(Cycle~1, \qref{O6})
\end{quote}

Questions also examined trustworthiness in the visual encodings. Trustworthiness is a value of qualitative research methods. For the Wordstreams, user's trust depended on being able to see how encodings were
derived and how they connected to the source text. We initially identified this as a preliminary theme of uncertainty. For instance, Expert~1 did not
know the inclusion cutoffs or \inquote{``what sudden attention or
TF-IDF mean''} (Cycle~1, \qref{O3}).
Similarly, Expert~2 found it
\inquote{``unclear to me how the visualization chooses which words to
include or exclude''} (Cycle~2, \qref{O3}).
They also requested features to \inquote{``customize or modify the topic/phrasing list''} (Cycle~2, \qref{O5}), suggesting that distrust, transparency and control were linked in participants' minds.

Notably, the member check did \emph{not} resolve the skepticism-enthusiasm into a single stance related to trust. 
Confirmatory item \qref{C4} (distrust when functions evoke uncertainty) was the weakest-endorsed of the five: among the six Cycle~2 respondents, only one agreed, one disagreed, and four were neutral, with several reframing what ``distrust'' meant. This weak endorsement explains why a preliminary 'uncertainty/confidence' theme was not fully explanatory. 
Participants redefined the concept of 'distrust' instead of confirming it, which highlighted the deeper disagreement reported in Theme 4. 
Expert~6 \textbf{tied trust to traceability}, \inquote{``Without seeing clear connections to the response text, I had a hard time trusting the visualizations''} (Cycle~2, \qref{C4}), whereas Expert~3  located the issue \textbf{not in uncertainty but in frequency bias itself} (Cycle~2, C4). 
The dissensus is therefore a meaningful disagreement about \emph{what the encoding is---and should be---for}, instead of a simple deviation from consensus. 
In other words, the visualization had hidden operations that evoked users' uncertainty and did not align with the values they associated with qualitative data.

\section{Discussion}
\label{sec:discussion}

\paragraph{Integration of findings.}
STEM instructor--researchers assessed education-facing tools through their own qualitative practice. 
The tools functioned as probes to elicit feedback about functionality and ideas about their deeper epistemological assumptions underlying the visual encoding.  
We found the Dashboard rates higher because it preserves in-context detail (\qref{L1}, \qref{L5}), addressing \hyperref[rq1]{RQ1}. 
The two tools are comparable for trend- and development-level reading (\qref{L2}, \qref{L6}).
However, the substantive finding (\hyperref[rq2]{RQ2}) is that perceived value, and therefore the likelihood of technology acceptance, was dependent on participants underlying epistemic values about the qualitative-quantitative approaches. 
In other words, some participant accepted a frequency-based view as a legitimate qualitative instrument. Others remained skeptical, considering them, at most, an entry point into qualitative analysis practices.

\subsection{Findings upon the dissensus}
The disagreement over frequency encoding's value (Theme 4) did not resolve, and we read this as informative to our pursuit rather than as a failure to reach consensus. It resonates with long-standing concerns about what aggregation and emphasis can hide~\cite{correll2019ethical} and about whose questions an encoding serves~\cite{dork2013critical}. The dissensus here may be due to competing role demands due to \textbf{our participants' multiple roles and perspectives}: as instructors making classroom decisions and as researchers committed to particular methodological traditions. Participants' comments often followed their research identity: a self-described grounded theorist valued the Dashboard's transparency (Expert 10, \qref{C1}), whereas a scholar attentive to under-represented voices raised concerns about frequency bias on equity grounds (Expert 3, \qref{O4}). They rated the distrust item (\qref{C4}) as the weakest-endorsed and several participants re-framed what ``distrust'' meant, relating it to uncertainty in the representation. 

\subsection{Implications}
Extensive research has established previously TAM as a reliable model to explain the adoption and actual use of educational technology ~\cite{davis1989perceived,SCHERER201913, or2024thirty}. However, the basic TAM model fell short of explaining the expert user's judgments of appropriateness for qualitative learning analytics. A conceptual model of TAM may need to consider not only the context of use associated with participants’ envisioned \textbf{role and task}, but also the \textbf{epistemic frames} that users bring.  An updated model to design visual platforms better is shown in Figure ~\ref{fig:tam-extension}).

Incorporating user roles and epistemic frames could help designers to consider the functions, methods, and values that are deeply debated and grounded in the qualitative research traditions. To distinguish roles, designers, evaluators and researchers could be define users within an institutional context, their professional identity, the functions they perform, or their credentials ~\cite{SCHERER201913}. 

\begin{figure}[!h]
    \centering
    \begin{tikzpicture}[
  font=\sffamily\large, >={Stealth[length=2.6mm]},
  node distance=1.0cm and 1.5cm,
box/.style={rectangle, 
    rounded corners, 
    draw=black, 
    thick, 
    align=center, 
    minimum width=3.5cm, 
    minimum height=1.5cm, 
    fill=gray!10
    },
main/.style={
    rectangle, 
    rounded corners, 
    draw=black, 
    thick, 
    align=center, 
    minimum width= 3cm, 
    minimum height=1.5cm, 
    fill=green!10 
    }, 
arrow/.style={ ->, thick }
]

\node[box] (ease) {\textbf{Perceived}\\\textbf{Ease of Use}}; 
\node[box,below=of ease] (useful) {\textbf{Perceived}\\\textbf{Usefulness}}; 
\node[main,right=of ease] (accept) {\textbf{Technology}\\\textbf{Acceptance}}; 

\node[box,left=of ease] (context) {\textbf{Context of Use}\\(Role \& Tasks)}; 
\node[box,above=of context] (epistemic) {\textbf{Epistemic Frame}\\(Qualitative Values)};

\draw[arrow] (useful) -- (accept); 
\draw[arrow] (ease) -- (accept); 

\draw[arrow] (context) -- (useful); 
\draw[arrow] (context) -- (ease); 
\draw[<->,dashed] (epistemic) -- (context);
\draw[arrow,dashed] (epistemic) -- (accept); 

\end{tikzpicture}   %
\caption{Proposed extension of the Technology Acceptance Model (TAM) for qualitative learning analytics. User's context of use (e.g., instructor, researcher) and epistemic frame (e.g., assumptions about valid qualitative practices) may affect the  perceived ease of use and usefulness. Solid arrows indicate established research relationships. Dashed arrows represent novel hypotheses.} 
    \label{fig:tam-extension} 
\end{figure}

However, investigating researcher perspectives and epistemic beliefs about qualitative data may require greater effort. This is because each method aligns with a sets of paradigms: positivism, post-positivism, critical theory, and constructivism ~\cite{guba2005paradigmatic}. These paradigms have come from different traditions and have deeply embedded assumptions. To add to the challenge, the distinctions among qualitative research paradigms may be implicit, contradictory, partially formed or still evolving through multidecadal debates ~\cite{debatereview,guba2005paradigmatic, lincoln2011paradigmatic}. 

Alternatively to understand  epistemic frames, researchers and designers may benefit from learning general practices that span multiple traditions of the qualitative research. These values include trustworthiness, authenticity, credibility, dependability, confirmability, plausibility, and transferability ~\cite{lincoln1995emerging, tracy2017big}. 
These values are evident in long standing qualitative reporting norms as well (e.g., thick descriptions, data crystallization, and multivocality) ~\cite{tracy2017big}. Thus, tools and functions that support in support users aligned with these values may be perceived as more valuable for qualitative data exploration than frequency-centered perspectives. We therefore aim to design with these components in mind, motivate our design recommendations.

\subsection{Design Recommendations for Qualitative Learning Analytics}
\label{sec:design_implications}

Consideration of the integrated results informs design implications for tools that bring quantitative
techniques into qualitative learning analytics (\hyperref[rq3]{RQ3}).  Below, we relate our recommendations to prior methodological arguments from qualitative research and text-analysis systems. 

\subsubsection{Enable custom data uploads with insight extraction}
\label{di:custom_data}

A unified tool should let users import their own corpus, adjust the Maker's display parameters while preserving the Dashboard's context retrieval path from any stream element back to the verbatim responses. This design suggestion aligns with previous work in supporting users to transition between aggregate and context-specific views ~\cite{text_voyant}. The priority of user control by increasing interactivity and access to detail comes with tradeoffs to representation, dashboard structure, and page layout simplicity \cite{dashboarddesign}.
 
\subsubsection{Make encoding choices transparent \& controllable}
\label{di:transparency}
Tools should (1) provide plain-language annotations explaining all thresholds and scoring methods, (2) allow users to adjust these thresholds and methods, and (3) offer an editable list of topics and phrases, with stop words removed automatically by default but adjustable by the user. These suggestions align with Chuang et al.~\cite{chuang2012interpretation}, who recommend exposing adjustable parameters and enabling progressive disclosure back to source data to make text visualizations interpretable and trustworthy. Our findings extend these principles to qualitative educational analysis by showing how educators valued such controls.

\subsubsection{Preserve rare but critical responses}
\label{di:Preserve}
A frequency encoding surfaces the common and can bury the infrequent, which participants flagged as a risk to validity. 
Tools intended for student data should offer complementary views that surface the long tail, e.g., a low-frequency or outlier panel, or a per-response distribution, so that diversity of voice is not lost to aggregation. 
Such features would support qualitative research values of credibility, trustworthiness, and confirmability ~\cite{lincoln1995emerging} and support critical research traditions ~\cite{ debatereview, guba2005paradigmatic, lincoln2011paradigmatic}. 
For instance in a thematic analysis, this design would realizes Braun and Clarke's distinction between thematic prevalence and analytic significance~\cite{braun2006using, braun2023toward} where a theme can be important to the research question without being frequent. 
We translate such methodological concerns from participants and the literature into a visualization requirement to keep infrequent responses accessible alongside dominant patterns. 
This argument supports why rare responses matter; however, it does not validate a particular solution such as an outlier panel. 
 
\subsubsection{Support flexible units of analysis}
\label{di:flexible}
Participants wanted to recompose the same corpus around units other than the whole class. Tools should let users pivot the corpus by student, by prompt, and by distinct-speaker count. 
Functions that support these groupings—e.g., by overlaying several terms on one timeline—would let one tool serve the several disciplinary uses named under Theme~3. 
In a recent work, ThemeClouds~\cite{colonel2025word} developed participant-level weighting in qualitative interview visualizations.
Our participants additionally requested regrouping by student, prompt, and conceptual category.
Such methods could be directly relevant to grounded theory methods and support values of data triangulation, transparency, and trustworthiness ~\cite{Interplay, debatereview}. Furthermore, the suggestion by Expert 5 to merge with pre-existing text analysis tools suggest that mixed-methods research may also benefit from a flexible unit of analysis.

\subsubsection{Use frequency as an entry point instead of endpoint}
\label{di:frequency}
We recommend designing the encoding explicitly as a bridge: the first step of a pipeline that provides one-click navigation from a stream element to its matching, relevant excerpts, plus export of selected words, phrases, and responses into coding workflows. This positioning is consistent with McNaught and Lam's~\cite{mcnaught2010using} characterization of word clouds as supplementary tools for preliminary analysis. As noted by Expert 1, 2, 7, and 10, designs that support early stages to get familiar with a data set would be extremely valuable and are often a first step for qualitative analyses and grounded theory ~\cite{braun2023toward,diehl2020studying}.
The tool can then align the encoding with how these researchers actually work, complementing existing qualitative analysis practices. This would help address the strongest objections raised in Theme~4.

\subsection{Limitations} The sample is small ($n{=}10$), chosen for dual expertise in qualitative methods and educational assessment, so the ratings are descriptive indicators reported without inferential tests~\cite{chen2021methods}.
This instrument is grounded in the constructs of technology acceptance but reliability and validity measures are needed to scale findings and applications of the assessment. 
Instead, we attempted triangulated findings through the open-ended and confirmatory items. Our claims are therefore interpretive, not generalizable. This also resonates with Losev et al. ~\cite{losev2022embracing}; scalability is not usually a goal in qualitative studies. Instead, we strive for transferability, so this model and design suggestions can be explored in other setting. The question designs should be revisited, recognizing that participants may envision the application in several roles as uncovered in our analysis of instructor-researchers. 

Limitations are also related to the WordStream platforms. We selected Dashboard and Maker due to our historic familiarity. The two tools are deliberately designed to be asymmetric, expanding the initial purpose-built design of the Dashboard into lightweight, web-based platform Maker. We used the asymmetry as probes under the assumption that it would give us insights into the stages of a workflow, not as matched competitors, so the comparison reflects context of use as much as the artifacts themselves. Readers should be cautioned from generalizing since  findings and recommendations come from a user study of two specialized platforms. 

All participants analyzed the same single corpus from one Data Visualization course; several anticipated that working with their own data would change their judgments. 
Finally, the data are self-reported responses, but do not investigate the fidelity between participant's stated and enacted practices. Given the variety of roles and qualitative methods, future work should consider case-specific uses, comparisons, and personas.

\section{Conclusion}
\label{sec:conclusion}
 
We set out to learn what qualitative education researchers can teach visualization designers, using two WordStream-based platforms: the Journal Data Dashboard and WordStream Maker, as probes into creating and reading visualizations of student journal writing for learning analytics. Ten STEM instructor-researchers rated the in-context Dashboard higher on perceived ease of use and usefulness, especially for exploring response detail, and were largely comparable on trend- and development-level reading. Alongside those ratings, our thematic analysis surfaced a sustained valuing of control and context (Themes~1--3) and, most importantly, an unresolved epistemological dissensus over whether a frequency-based encoding can meaningfully represent qualitative data (Theme~4). These findings support an argument that (1) perceived ease of use, (2) perceived usefulness, (3) context of use or role, and (4) epistemic frames are relevant to users of qualitative data visualization and learning analytic tools.

We translated these findings into five design implications: (1) unify creation and extraction; (2) make encoding transparent and controllable; (3) preserve rare but critical responses; (4) support flexible units of analysis; and (5) treat frequency as an entry point rather than an endpoint.
These principles are an essential step to integrate quantitative technique with qualitative learning analytics inquiry. We hope these recommendations inform tools that let instructors and researchers move flexibly between the shape of a cohort's language and the individual voices that make it up.

\acknowledgments{
This work was supported in part by the National Aeronautics and Space Administration (NASA) and the Gordon Research Conferences under NASA Award \#17-TWSC17-0055. This work was also supported in part by the National Nuclear Security Administration (NNSA) Advanced Simulation and Computing (ASC) Program under grant number DE-SCL0000089. This work is released under LA-UR-26-29665.

The authors would like to thank Riyansha Goyal (RG), Jaime Serrano (JS), and Jayrylle Jaylo (JJ) for their support of this study. Funding support for research students came from the Scholarship, Research, and Creative Practice (SRCP) Seed Award and School of Interdisciplinary Arts and Sciences at the University of Washington Bothell. }

\bibliographystyle{abbrv-doi}

\bibliography{references}
\end{document}